# Quantum Critical Transition and Kondo Screening of Magnetic Moments in Graphene

Jinhai Mao[1,†], Yuhang Jiang[1,†], Po-Wei Lo[2,3,*], Daniel May[4], Guohong Li[1], Guang-Yu Guo[2,3], Frithjof B. Anders[4], Takashi Taniguchi[5], Kenji Watanabe[5] and Eva Y. Andrei[1]

[1]Rutgers University, Department of Physics and Astronomy, 136 Frelinghuysen Road, Piscataway, NJ 08855 USA

[2]Department of Physics, National Taiwan University, Taipei 10617, Taiwan

[3]Physics Division, National Center for Theoretical Sciences, Hsinchu 30013, Taiwan

[4]Theoretische Physik 2, Technische Universität Dortmund, 44221 Dortmund, Germany

[5]Advanced Materials Laboratory, National Institute for Materials Science, 1-1 Namiki, Tsukuba 305-0044, Japan

[†]These authors contributed equally

In normal metals, the magnetic-moment of impurity-spins disappears below a characteristic "Kondo" temperature, $T_K$, where coupling with the conduction-band electrons produces an entangled state that screens the local moment(*1*). In contrast, moments embedded in insulators remain unscreened at all temperatures. This raises the question about the fate of magnetic-moments in intermediate, pseudogap systems, such as graphene. In these systems theory predicts a quantum phase-transition at a critical coupling strength which separates a local magnetic-moment phase from a Kondo-screened phase(*2-7*) as illustrated in Fig. 1A. However, attempts to experimentally confirm these predictions and their intriguing consequences such as the ability to electrostatically control magnetic-moments, have thus far been elusive(*8-10*). Here we report the observation of Kondo screening and the quantum phase-transition between screened and unscreened phases of vacancy magnetic-moments in graphene. Using scanning-tunneling-microscopy (STM), spectroscopy (STS) and numerical-renormalization-group (NRG) calculations, we identified Kondo-screening by its spectroscopic signature and mapped the phase-transition as a function of coupling strength and chemical potential. We show that this transition makes it possible to turn the magnetic-moment on and off electrostatically through a gate voltage or mechanically through variations in local curvature.

Graphene, with its tunable chemical potential(*11, 12*), provides a playground for exploring the physics of the magnetic quantum phase-transition. But embedding a magnetic-moment and producing sufficiently large coupling with the itinerant electrons in graphene, poses significant experimental challenges: adatoms typically reside far above the graphene plane, while substitutional atoms tend to become delocalized and non-magnetic (*13*). An alternative and efficient way to embed a magnetic-moment in graphene is to create single atom vacancies. The removal of a carbon atom from the honeycomb lattice induces a magnetic moment stemming from the unpaired electrons at the vacancy site (*14-17*) . This moment has two contributions: one is a resonant state (zero mode - ZM) at the Dirac point (DP) due to the removal of an electron from the $\pi$ band; the other arises from the broken $\sigma$-orbitals, two of which hybridize leaving a dangling bond that hosts an unpaired electron (*17*). The ZM couples ferromagnetically to the dangling $\sigma$ orbital (*17*) as well as to the conduction electrons (*18, 19*) and remains unscreened. In flat graphene the magnetic moment from the dangling $\sigma$-bond is similarly unscreened because the $\sigma$ orbital is orthogonal to the $\pi-$ band conduction electrons (*19, 20*). However, it has been proposed that this constraint would be eliminated in the presence of a local curvature which removes the orthogonality of the $\sigma$ orbital with the conduction band, and enables Kondo screening(*19, 21*).

We employed STS (*22, 23*) to identify Kondo-screening of the vacancy magnetic-moment by the distinctive zero-bias resonance it produces in the dI/dV curves (I is the tunneling current and V the junction bias), hereafter called "Kondo peak". We first discuss samples consisting of two stacked single layer graphene sheets on a $SiO_2$ substrate (G/G/$SiO_2$) capping a doped Si gate electrode (Fig. 1B). A large twist angle between the two layers ensures electronic decoupling, and preserves the electronic structure of single layer graphene while reducing substrate induced random potential fluctuations (*24, 25*). A further check of the Landau-level spectra in a magnetic field revealed the characteristic sequence expected for massless Dirac fermions (*11, 22*), confirming the electronic decoupling of the two layers (SI).Vacancies were created by low energy (100 eV) $He^+$ ion sputtering followed by in situ annealing (*8, 12, 26*). In STM topography of a typical irradiated sample (Fig. 1C) the vacancies appear as small protrusions on top of large background corrugations. To establish the nature of a vacancy we zoom in to obtain atomic resolution topography and spectroscopy. Single atom vacancies are recognized by their distinctive triangular $\sqrt{3} \times \sqrt{3}\, R30°$ topographic fingerprint (Fig. 1C inset) (*12, 26, 27*) which is accompanied by a pronounced peak in the dI/dV spectra at the DP reflecting the presence of the ZM. If both these features are present we identify the vacancy as a single atom vacancy (See SI for other types of defects) and proceed to study it further. In order to separate the physics at the DP and the Kondo screening which produces a peak near $E_F \equiv 0$, the spectrum of the vacancy in Fig. 1D is taken at finite doping corresponding to a chemical potential, $\mu \equiv E_F - E_D = -54\ meV$. Far from the vacancy (lower curve), we observe the 'V' shaped spectrum characteristic of pristine graphene, with the minimum identifying the DP energy. In contrast, at the center of the vacancy (Fig. 1D upper curve), the spectrum features two peaks, one at the DP identifying the ZM and the other at zero bias corresponding to the Kondo peak (*1*). (In STS the zero-bias is identified with the Fermi energy, $E_F$.) From the line-shape of the Kondo peak (Fig. 1F inset), we extract $T_K = (67 \pm 2)\ K$ by fitting to the Fano line-shape (*28, 29*) characteristic of Kondo resonances (SI, Section 3). As a further independent check we compare in Fig. 1F the temperature dependence of the linewidth to that expected for a Kondo-screened impurity(*28, 30*), $\Gamma_{LW} = \sqrt{(\alpha k_B T)^2 + (2k_B T_K)^2}$ from which we obtain $T_K = (68 \pm 2)K$, consistent with the above value, and α = 6.0 ± 0.3 in agreement with measurements and numerical simulations on ad-atoms (*28, 31*). Importantly, as we show below, this resonance is pinned to $E_F$ over the entire range of the chemical potential, as expected for the Kondo peak (*32, 33*).

The gate dependence of the spectra corresponding to the hundreds of vacancies studied here falls into two clearly defined categories, which we label type *I*, and type *II*. In Fig. 2A we show the evolution with chemical potential of the spectra at the center of a type *I* vacancy. Deep in the p-doped regime, we observe a peak which is tightly pinned to, $E_F$, consistent with Kondo-screening. Upon approaching charge neutrality the Kondo peak disappears for $\mu \geq$ -58 and reenters asymmetrically in the n-doped sector, for $\mu \geq 10 meV$. As we discuss below, the absence of screening close to the charge neutrality point and its reentrance in the n-doped regime for type I vacancies is indicative of pseudogap Kondo physics for subcritical coupling strengths (*6, 34*). For type II vacancies, the evolution of the spectra with chemical potential, shown in Fig. 2B, is qualitatively different. The Kondo peak is observed in the p-doped regime and disappears close to charge neutrality, but does not reappear on the n-doped side. We show below that this behavior is characteristic of pseudogap Kondo physics for vacancies whose coupling to the conduction band is supercritical (*6, 34*).

To better understand these results we performed NRG calculations for a minimal model based on the pseudogap asymmetric Anderson impurity model (*4, 19, 35*) comprising the free local σ-orbital coupled to the itinerant π-band. In the Anderson model, which is relevant to our case, the impurity magnetic moment is finite only for $\varepsilon_d \leq E_F < \varepsilon_d + U$ and vanishes outside this range. Here $\varepsilon_d$ is the impurity energy and U is the Coulomb interaction strength. For $E_F \geq \varepsilon_d + U$ the magnetic moment vanishes due to the singlet formed by double occupancy rather than due to Kondo screening, hence the term "frozen impurity". This scenario stands in contrast to the standard Kondo model where the bare impurity moment is independent of $E_F$. Using $\varepsilon_d = -1.6\ eV$ for the bare σ-orbital energy (*14, 34*) and $U = 2$ eV (*14, 36, 37*) in the p-doped regime (see SI Section 7 for n-doped) we calculated the spectral function for several values of the coupling strength, $\Gamma_0$, (SI, Section 8) from which we determined the critical coupling $\Gamma_c$ = 1.15 eV that separates the local-magnetic-moment and the frozen-impurity phases at μ = 0. Using the NRG calculation to fit the measured STS spectra we obtained the value of the reduced coupling $\Gamma_0/\Gamma_C$ for each vacancy in Fig. 3 (SI, Section 9). The value, $\Gamma_0/\Gamma_C$ = 0.90, and 1.83 obtained for the spectra in Fig. 2A and Fig. 2B places these two vacancies in the sub-critical and super-critical regimes respectively.

In Fig. 3 we compare the chemical-potential dependence of the measured $T_K$, with the numerical simulation results. The $T_K$ values are obtained from Fano-fits of the Kondo peaks leading to the $T_K$ (μ) curves, shown in Fig. 3A. The corresponding values of $\Gamma_0/\Gamma_C$ and the $T_K$ (μ) curves obtained by using NRG to simulate the spectra are shown in Fig. 3B. The close agreement between experiment and simulations confirms the underlying pseudogap Kondo physics. In Fig. 4A we summarize the numerical results in a μ-$\Gamma_0$ phase diagram. At charge neutrality (defined by the μ = 0 line), the critical point $\Gamma_0/\Gamma_C$ =1 signals a quantum phase transition between the Local-Magnetic-Moment phase and the Frozen-Impurity phase (*34*). The Kondo-Screened phase appears at finite doping (μ ≠ 0) and is marked by the appearance of the Kondo-peak, (*6, 7*). The phase diagram clearly shows a strong electron-hole asymmetry consistent with the asymmetric Kondo screening expected in pseudogap systems (*2*).

The single orbital model employed in the NRG simulations gives an accurate description of the experiment in the p-doped regime where the ZM is sufficiently far from the Kondo peak so that their interaction is negligibly small. This is evident from the agreement between experiment and simulation in this regime (Fig. 3). Upon approaching charge neutrality, interactions between the two orbitals through Hund's coupling and level repulsion become relevant. As described in the SI we introduced an effective Coulomb interaction term to take into account this additional repulsion. Preliminary results for a comprehensive two-orbital pseudogap Anderson impurity model, which will be published elsewhere, indicate that the single orbital model together with the phenomenological correction accounting for the enhanced Coulomb interaction when the ZM is occupied captures the main features of the Kondo physics reported here.

Theoretical work (*19, 38*) suggests that finite Kondo screening of the vacancy moment may occur if corrugations in the graphene layer produce an out of plane component of the dangling σ orbital. This removes the orthogonality restriction(*20*) that prevents hybridization of the σ and π bands in flat graphene and produces a finite coupling-strength which increases monotonically with the out of plane projection of the orbital(*21, 39, 40*). To check this conjecture we repeated the experiments for samples on substrates with different average corrugation amplitudes as shown in Fig. 4C. For consistency all the fabrication steps were identical. In the G/SiO$_2$ sample (single layer graphene on SiO$_2$) where the corrugation amplitude was largest (~1nm), 60% of the vacancies displayed the Kondo peak and $T_K$ attained values as high as 180K (SI, Section 5). For the flatter G/G/SiO$_2$ where the average corrugation was ~0.5nm, we found that 30% of the vacancies showed the Kondo peak. For samples deposited on hBN, which were the flattest with local corrugation amplitudes of ~0.1nm, none of the vacancies showed the Kondo peak(*12*). This is illustrated in Fig. 1E showing a typical dI/dV curve on a vacancy in G/G/BN (double layer graphene on hBN) where a gate voltage of Vg = -30V was applied to separate the energies of $E_F$ and the DP. While this spectrum shows a clear ZM peak, the Kondo peak is absent over the entire range of doping(*12*). The absence of the Kondo peak in all the samples deposited on hBN highlights the importance of the local curvature. In order to quantify the effect of the local curvature on the coupling strength, we employed STM topography to measure the local radius of curvature, R, at the vacancy sites ( Fig. 4D inset) from which we estimate the angle between the σ orbital and the local graphene plane orientation (*39*), $\theta \approx a/2R$, where *a* is the lattice spacing. We find that the coupling strength, $\Gamma_0/\Gamma_C$ (θ), shows a monotonic increase with θ (Fig. 4D), consistent with the theoretical expectations (*19, 21, 39*). Interestingly, the effect of the curvature on the Kondo coupling was also observed for Co atoms deposited on corrugated graphene (*41*), and was also utilized to enhance the spin-orbit coupling (*42*).

These results shed new light on the contradictory conclusions drawn from earlier magnetometry (*9, 10*) and transport measurements(*8*) on irradiation induced vacancies in graphene. While the transport measurements revealed a resistivity minimum and logarithmic scaling indicative of Kondo screening with unusually large values of $T_K$ ~ 90K, magnetometry measurements showed Curie behavior with no evidence of low-temperature saturation, suggesting that that the vacancy moments remained unscreened. To understand the origin of this discrepancy we note that magnetometry and transport are sensitive to complementary aspects of the Kondo effect. The former probes the magnetic moment and therefore only sees vacancies that are not screened, while the latter probes the enhanced scattering from the Kondo cloud which selects only the vacancies whose moment is Kondo screened. Importantly, these techniques take a global average over all the vacancies in the sample. This does not pose a problem when all the impurities have identical coupling strengths. But if there is a distribution of couplings ranging from zero to finite values, as is the case here, global magnetization and transport measurements will necessarily lead to opposite conclusions as reported in the earlier work.

The local spectroscopy technique employed here made it possible to disentangle the physics of Kondo screening in the presence of a distribution of coupling strengths. This work demonstrates the existence of Kondo screening in a pseudogap system and identifies the quantum phase transition between a screened and an unscreened local magnetic moment. It further shows that the local magnetic moment can be controlled both electrically and mechanically, by using a gate voltage and a local curvature respectively.

Acknowledgments: We acknowledge support from DOE-FG02-99ER45742 (E.Y.A. and J.M), NSF DMR 1708158 (Y.J.), Ministry of Science and Technology and also Academia Sinica of Taiwan (G.Y.G. and P.W.L.), Deutsche Forschungsgemeinschaft via project AN-275/8-1 (D.M and F.B.A).

* Current address: Department of Physics, Cornell University, Ithaca, New York 14853

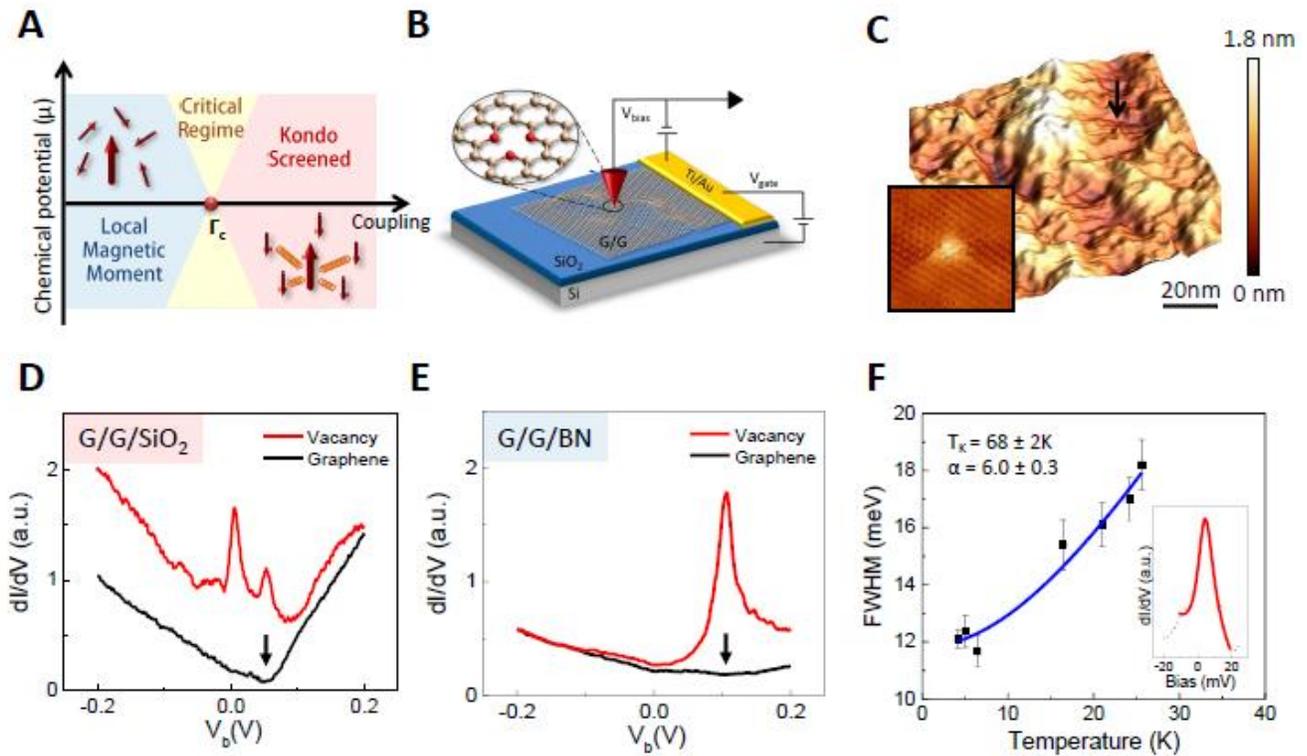

**Fig 1. Kondo peak at a single-atom vacancy in graphene.** (**A**) Schematic phase diagram of the pseudo-gap Kondo effect. The critical regime (yellow) separates the Local-magnetic-moment phase from the Kondo-screened phase. Arrows represent the ground state of the system with the large arrow corresponding to the local spin and the smaller ones representing the spins of electrons in the conduction band. (**B**) Schematics of the experimental setup. (**C**) STM topography of a double layer graphene on $SiO_2$ (G/G/$SiO_2$). The arrow indicates an isolated vacancy ($V_b$ = -300mV, $I$ = 20pA, $V_g$ = 50V). Inset: atomic resolution topography of a single atom vacancy shows the distinctive triangular structure (4nm x 4nm), $V_b$ = -200mV, $I$ = 20pA, $V_g$ = -27V. (**D**) $dI/dV$ spectra at the center of a single atom vacancy (upper red curve) and on pristine graphene far from the vacancy (lower black curve). The curves are vertically displaced for clarity ($V_b$ = -200mV, $I$ = 20pA, $V_g$ = 0V). The arrow labels the Dirac point. (**E**) Same as (D) but for a vacancy in a G/G/BN sample ($V_b$ = -200mV, $I$ = 20pA, $V_g$ = -30V). (**F**) Evolution of the measured linewidth with temperature (black data points) shown together with the fit (blue solid line) discussed in the text. Inset: Zoom into the Kondo peak (black dotted line) together with the Fano lineshape fit (red solid line).

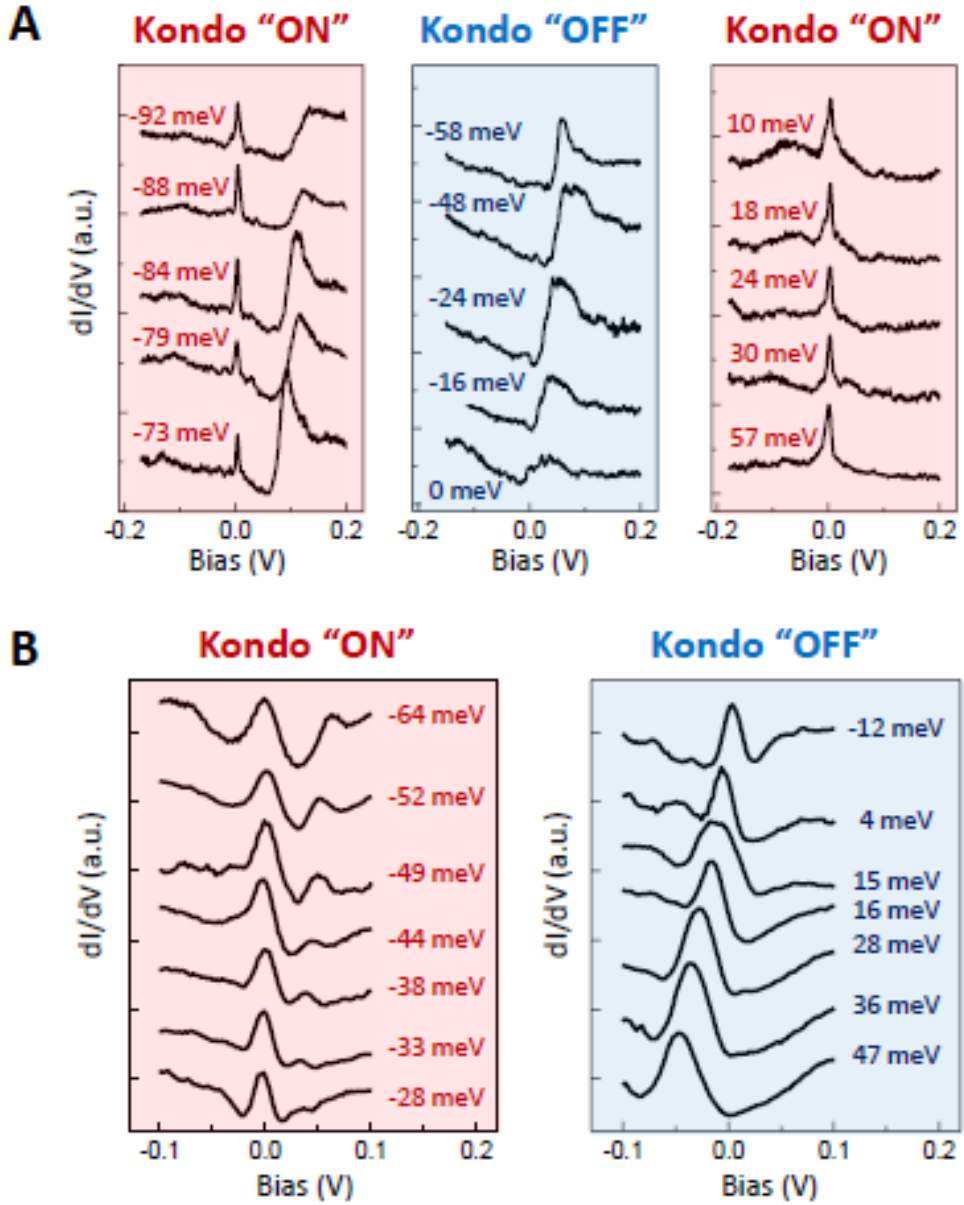

**Fig. 2 Evolution of Kondo screening with chemical potential.** (**A**) dI/dV curves for a subcritical Kondo vacancy (type I in text) with reduced coupling strength $\Gamma_0/\Gamma_C = 0.90$ at the indicated values of chemical potential. Red (blue) shade indicates the presence (absence) of the Kondo peak ($V_b = -200$mV, $I = 20$pA). The chemical potential is tuned by the backgate voltage (*12*). (**B**) dI/dV curves for a supercritical Kondo vacancy (type II in text) with $\Gamma_0/\Gamma_C = 1.83$.

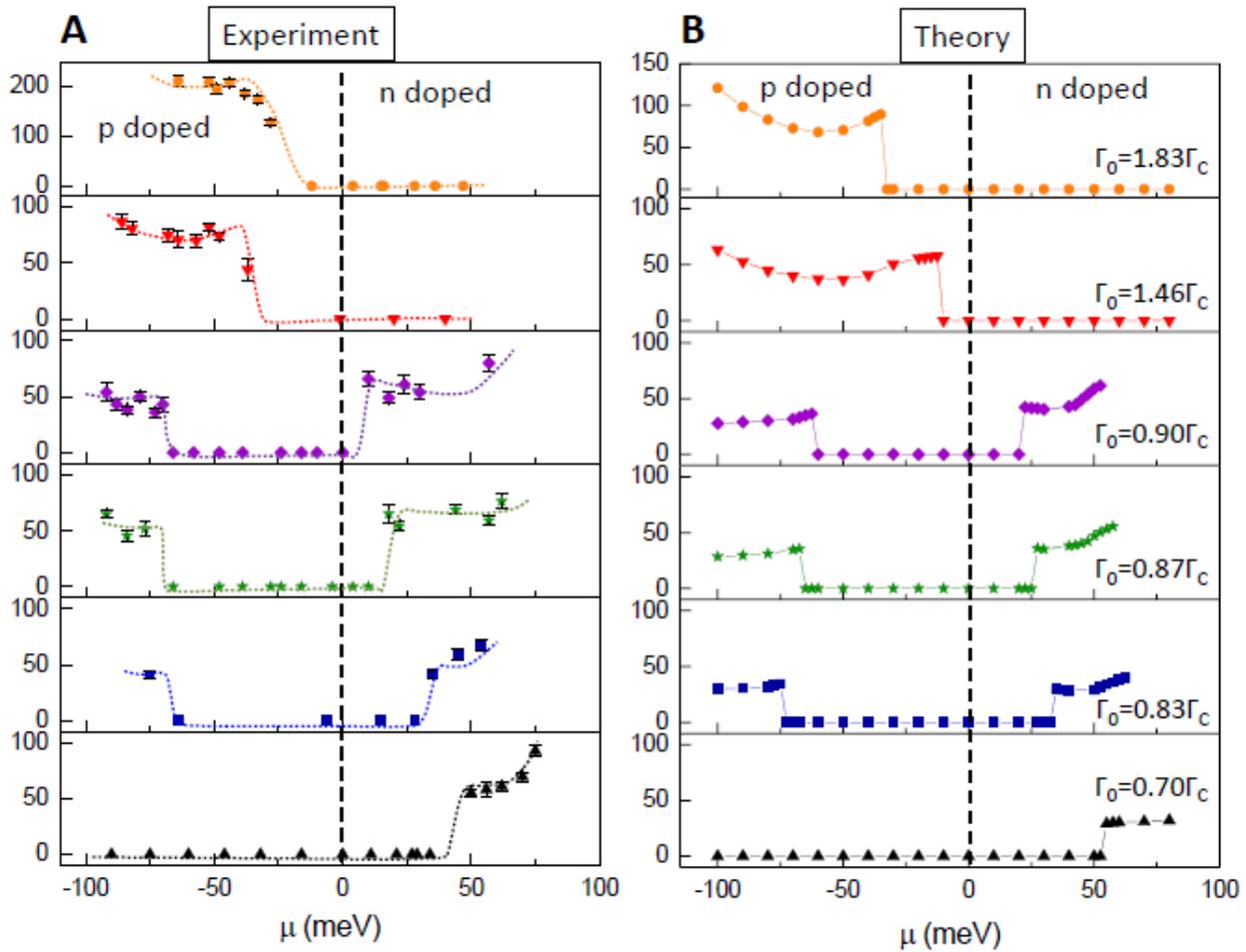

**Fig. 3 Chemical-potential dependence of the Kondo temperature**. (A) Chemical potential dependence of $T_K$ obtained from the Fano lineshape fit of the Kondo peak. In the regions where the peak is absent we designated $T_K = 0$. $\Gamma_0/\Gamma_C$ is obtained by comparing to the NRG results in B. (B) NRG result for the vacancies in panel A. $T_K$ is estimated by fitting the numerically simulated Kondo peak (SI).

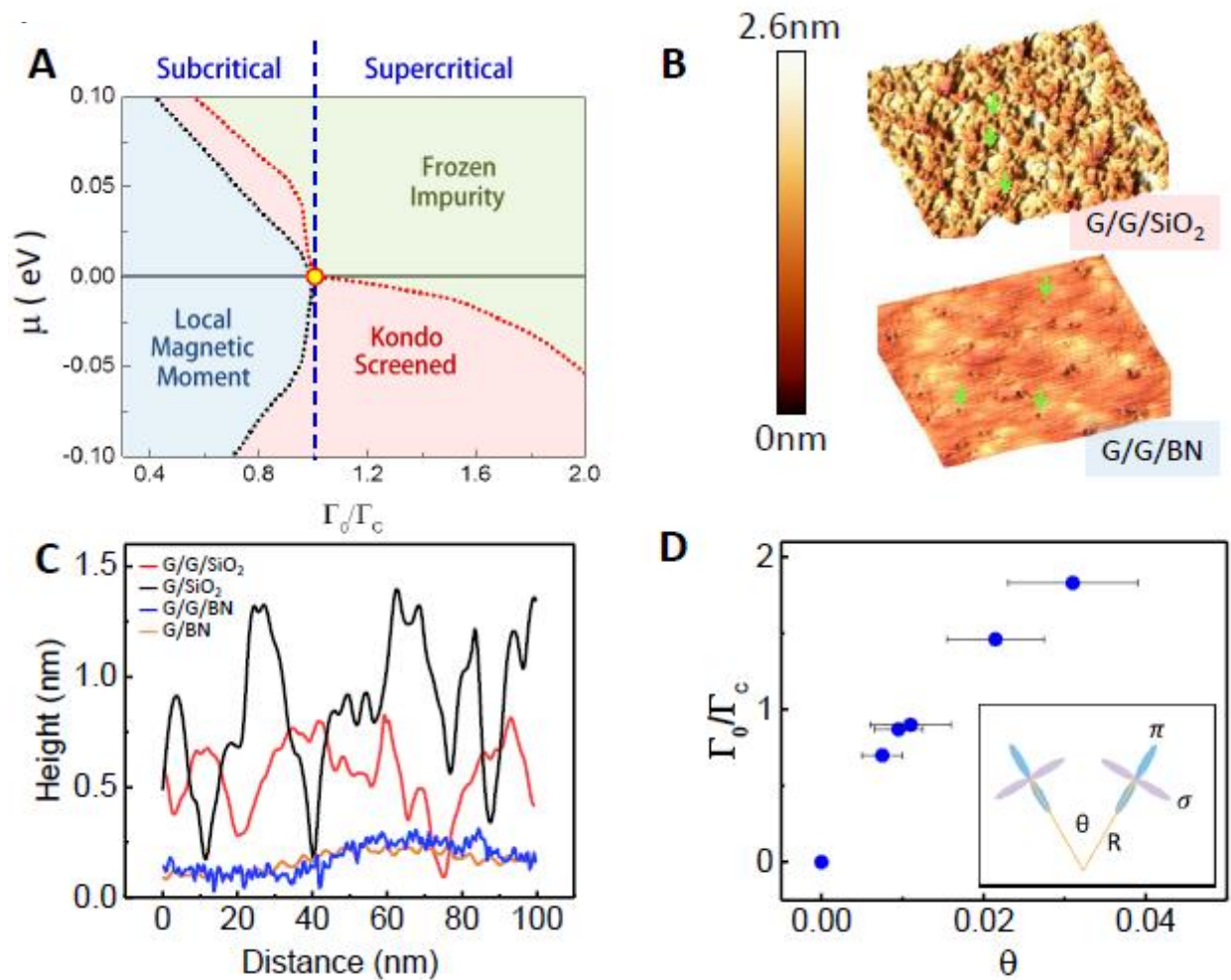

**Fig. 4 Quantum phase-transition and Kondo screening.** (**A**) $T_K$−$\Gamma_0$ phase diagram at 4.2K. The critical coupling $\Gamma_C$ (circle at $\Gamma_0/\Gamma_C =1.0$ ) designates the boundary between Frozen-Impurity and the Local-Magnetic-Moment phases at $\mu$=0. Dashed lines represent boundaries between the different phases. (**B**) STM topography for the G/G/SiO$_2$ (top) and G/G/BN (bottom) samples with the same scale bar ($V_b$ = -300mV, $I$ = 20pA). The arrows point to the vacancies. (**C**) Typical line profile of the STM topographies of graphene on different substrates with the same scanning parameters as in B. (**D**) The evolution of the hybridization strength with the curvature. Inset shows a geometrical sketch of curvature effect on the orbital hybridization.

References:

1. A. C. Hewson, *The Kondo Problem to Heavy Fermions*. *(Cambridge University Press, Cambridge)*, (1997).
2. D. Withoff, E. Fradkin, Phase-Transitions in Gapless Fermi Systems with Magnetic-Impurities. *Phys Rev Lett* **64**, 1835 (1990).
3. C. R. Cassanello, E. Fradkin, Kondo effect in flux phases. *Phys Rev B* **53**, 15079 (1996).
4. C. Gonzalez-Buxton, K. Ingersent, Renormalization-group study of Anderson and Kondo impurities in gapless Fermi systems. *Phys Rev B* **57**, 14254 (1998).
5. L. Fritz, M. Vojta, Phase transitions in the pseudogap Anderson and Kondo models: Critical dimensions, renormalization group, and local-moment criticality. *Phys Rev B* **70**, 214427 (2004).
6. M. Vojta, L. Fritz, R. Bulla, Gate-controlled Kondo screening in graphene: Quantum criticality and electron-hole asymmetry. *Epl-Europhys Lett* **90**, 27006 (2010).
7. T. Kanao, H. Matsuura, M. Ogata, Theory of Defect-Induced Kondo Effect in Graphene: Numerical Renormalization Group Study. *J Phys Soc Jpn* **81**, 063709 (2012).
8. J. H. Chen, L. Li, W. G. Cullen, E. D. Williams, M. S. Fuhrer, Tunable Kondo effect in graphene with defects. *Nat Phys* **7**, 535 (2011).
9. R. R. Nair *et al.*, Spin-half paramagnetism in graphene induced by point defects. *Nat Phys* **8**, 199 (2012).
10. R. R. Nair *et al.*, Dual origin of defect magnetism in graphene and its reversible switching by molecular doping. *Nat Commun* **4**, 3010 (2013).
11. E. Y. Andrei, G. H. Li, X. Du, Electronic properties of graphene: a perspective from scanning tunneling microscopy and magnetotransport. *Rep Prog Phys* **75**, 056501 (May, 2012).
12. J. H. Mao *et al.*, Realization of a tunable artificial atom at a supercritically charged vacancy in graphene. *Nat Phys* **12**, 545 (2016).
13. H. Wang *et al.*, Doping Monolayer Graphene with Single Atom Substitutions. *Nano Letters* **12**, 141 (2012).
14. O. V. Yazyev, L. Helm, Defect-induced magnetism in graphene. *Phys Rev B* **75**, 125408 (2007).
15. B. Uchoa, V. N. Kotov, N. M. R. Peres, A. H. C. Neto, Localized magnetic states in graphene. *Phys Rev Lett* **101**, 026805 (2008).
16. J. J. Palacios, J. Fernandez-Rossier, L. Brey, Vacancy-induced magnetism in graphene and graphene ribbons. *Phys Rev B* **77**, 195428 (2008).
17. B. R. K. Nanda, M. Sherafati, Z. S. Popović, S. Satpathy, Electronic structure of the substitutional vacancy in graphene: density-functional and Green's function studies. *New J Phys* **14**, 083004 (2012).
18. P. Haase, S. Fuchs, T. Pruschke, H. Ochoa, F. Guinea, Magnetic moments and Kondo effect near vacancies and resonant scatterers in graphene. *Phys Rev B* **83**, 241408 (2011).
19. M. A. Cazalilla, A. Iucci, F. Guinea, A. H. Castro-Neto, Local Moment Formation and Kondo Effect in Defective Graphene. *arXiv:1207.3135 (2012).*
20. M. Hentschel, F. Guinea, Orthogonality catastrophe and Kondo effect in graphene. *Phys Rev B* **76**, 115407 (2007).
21. T. Ando, Spin-orbit interaction in carbon nanotubes. *J Phys Soc Jpn* **69**, 1757 (2000).
22. A. Luican, G. Li, E. Y. Andrei, Scanning tunneling microscopy and spectroscopy of graphene layers on graphite. *Solid State Communications* **149**, 1151 (2009).
23. G. Li, A. Luican, E. Y. Andrei, Self-navigation of an STM tip toward a micron sized sample. *Rev. Sci. Instrum.* **82**, 073501 (2011).

24. G. Li *et al.*, Observation of Van Hove singularities in twisted graphene layers. *Nat Phys* **6**, 109 (2010).
25. A. Luican *et al.*, Single-Layer Behavior and Its Breakdown in Twisted Graphene Layers. *Phys Rev Lett* **106**, 126802 (2011).
26. M. M. Ugeda, I. Brihuega, F. Guinea, J. M. Gómez-Rodríguez, Missing Atom as a Source of Carbon Magnetism. *Phys Rev Lett* **104**, 096804 (2010).
27. K. F. Kelly, D. Sarkar, G. D. Hale, S. J. Oldenburg, N. J. Halas, Threefold Electron Scattering on Graphite Observed with C60-Adsorbed STM Tips. *Science* **273**, 1371 (1996).
28. M. Ternes, A. J. Heinrich, W. D. Schneider, Spectroscopic manifestations of the Kondo effect on single adatoms. *J Phys-Condens Mat* **21**, 053001 (2009).
29. A. Schiller, S. Hershfield, Theory of scanning tunneling spectroscopy of a magnetic adatom on a metallic surface. *Phys Rev B* **61**, 9036 (2000).
30. K. Nagaoka, T. Jamneala, M. Grobis, M. F. Crommie, Temperature dependence of a single Kondo impurity. *Phys Rev Lett* **88**, 077205 (2002).
31. A. F. Otte *et al.*, The role of magnetic anisotropy in the Kondo effect. *Nat Phys* **4**, 847 (2008).
32. S. M. Cronenwett, T. H. Oosterkamp, L. P. Kouwenhoven, A tunable Kondo effect in quantum dots. *Science* **281**, 540 (1998).
33. D. Goldhaber-Gordon *et al.*, Kondo effect in a single-electron transistor. *Nature* **391**, 156 (1998).
34. P. W. Lo, G. Y. Guo, F. B. Anders, Gate-tunable Kondo resistivity and dephasing rate in graphene studied by numerical renormalization group calculations. *Phys Rev B* **89**, 195424 (2014).
35. D. A. Ruiz-Tijerina, L. G. G. V. Dias da Silva, Transport signatures of Kondo physics and quantum criticality in graphene with magnetic impurities. *Phys Rev B* **95**, 115408 (2017).
36. H. Padmanabhan, B. R. K. Nanda, Intertwined lattice deformation and magnetism in monovacancy graphene. *Phys Rev B* **93**, 165403 (Apr 4, 2016).
37. V. G. Miranda, L. G. G. V. Dias da Silva, C. H. Lewenkopf, Coulomb charging energy of vacancy-induced states in graphene. *Phys Rev B* **94**, 075114 (2016).
38. A. K. Mitchell, L. Fritz, Kondo effect with diverging hybridization: Possible realization in graphene with vacancies. *Phys Rev B* **88**, 075104 (2013).
39. D. Huertas-Hernando, F. Guinea, A. Brataas, Spin-orbit coupling in curved graphene, fullerenes, nanotubes, and nanotube caps. *Phys Rev B* **74**, 155426 (2006).
40. A. H. Castro Neto, F. Guinea, Impurity-Induced Spin-Orbit Coupling in Graphene. *Phys Rev Lett* **103**, 026804 (2009).
41. J. Ren *et al.*, Kondo Effect of Cobalt Adatoms on a Graphene Monolayer Controlled by Substrate-Induced Ripples. *Nano Letters* **14**, 4011 (2014/07/09, 2014).
42. J. Balakrishnan, G. Kok Wai Koon, M. Jaiswal, A. H. Castro Neto, B. Ozyilmaz, Colossal enhancement of spin-orbit coupling in weakly hydrogenated graphene. *Nat Phys* **9**, 284 (2013).

# Supplementary Information

# Quantum Critical Transition and Kondo Screening of Magnetic Moments in Graphene

Jinhai Mao[1†], Yuhang Jiang[1†], Po-Wei Lo[2,3], Daniel May[4], Guohong Li[1], Guang-Yu Guo[2,3], Frithjof Anders[4], Takashi Taniguchi[5], Kenji Watanabe[5] and Eva Y. Andrei[1]

[1]Rutgers University, Department of Physics and Astronomy, 136 Frelinghuysen Road, Piscataway, NJ 08855 USA

[2]Department of Physics, National Taiwan University, Taipei 10617, Taiwan

[3]Physics Division, National Center for Theoretical Sciences, Hsinchu 30013, Taiwan

[4]Theoretische Physik 2, Technische Universität Dortmund, 44221 Dortmund, Germany

[5]National Institute for Materials Science, 1-1 Namiki, Tsukuba 305-0044, Japan

[†]These authors equally contributed to the work

1. **Methods.**

G/G/SiO$_2$ sample used in the experiment is stacked two single layers graphene on 300nm SiO$_2$ dielectric layer capped highly doped Si chip (acting as the backgate electrode)(*1-3*). The first layer graphene is exfoliated to SiO$_2$ surface and the second layer graphene is stacked on top by dry transfer process. PMMA and PVA thin film is used as the carrier in the dry transfer process. Au/Ti electrodes are added by the standard SEM lithography, and followed by the metal thermal deposition process. After liftoff, the sample is annealed in forming gas (H$_2$ : Ar, 1:9) at 300℃ for 3 hours to remove the PMMA residue, and further annealed overnight at 230 ℃ in UHV(*1*). To introduce single vacancies in the graphene lattice, the device is exposed under UHV conditions to a beam of He$^+$ ions with energy 100eV for 5 to 10 seconds, and further annealed at high temperature in situ(*4*). Except where mentioned all the STM experiments are performed at 4.2K. dI/dV curves are collected by the standard lock-in technique, with 0.5mV AC modulation at 473Hz added to the DC sample bias(*5-7*). The chemical potential is tuned by the backgate voltage as illustrated shown in Fig.1A of the main text.

**2. Electronically decoupled top layer graphene.**

The two graphene layers were stacked with a large twist angle between them to ensure electronic decoupling (*3, 8*). This absence of moiré patterns for these samples is consistent with their decoupling. The single layer nature of the top layer is further revealed by the characteristic Landau level (LL) sequence of peaks, $E_N = E_D \pm \frac{\hbar v_F}{l_B}\sqrt{2|N|}$, $N = 0, \pm 1, ..$ which appears in the presence of a magnetic field. Here $E_D$ is the Dirac point (DP) energy measured relative to $E_F$, $v_F$ is the Fermi velocity and $l_B$ the magnetic length. Fig.1S shows the LL spectrum on the graphene surface far from any vacancy from which we extract the Fermi velocity $v_F = 1.04 \times 10^6$ m/s by fitting to the LL sequence. The specific LLs sequence as the fingerprint of massless Dirac fermions is the direct proof of that the two graphene layer are electronically decoupled.

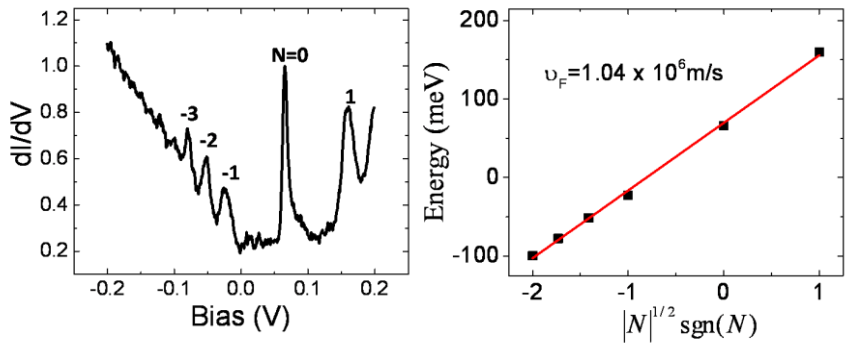

**Fig.1S**. (Left) Landau levels spectroscopy (LLs) of graphene under the magnetic field B = 6T. The numbers label the LL index. (Right) Fit the LLs sequence to $E_N = E_D \pm \frac{\hbar v_F}{l_B}\sqrt{2|N|}$, $N = 0, \pm 1, ..$ to extract the Fermi velocity.

## 3. Fano fitting to the Kondo peak.

The Kondo temperature is obtained by fitting the dI/dV curve to the Fano lineshape $\frac{dI}{dV} \propto \frac{(\varepsilon+q)^2}{1+\varepsilon^2} + A$ , where A is the background tunneling signal and q is the Fano asymmetry factor given by $q \propto t2/t1$ (t1 and t2 are the matrix elements for electron tunneling into the continuum of the bulk states and the discrete Kondo resonance, respectively)(9). Here $\varepsilon = \frac{eV - \varepsilon_0}{\Gamma/2}$ is the normalized energy ($\varepsilon_0$ is the position of the resonance and Γ is the full width of half maximum FWHM of the Kondo peak, which is related to Kondo temperature $T_K$ as $k_B T_K = \Gamma/2$). Due to full self-energy of the resonant level interactions and coupling to the conduction electrons, $\varepsilon_0$ may slightly shift away from $E_F$(10).

## 4. Temperature and position dependence of the Kondo peak.

The temperature dependence of the Kondo peaks is analyzed by raising the temperature from 4.2K to 25K, Fig.2S.

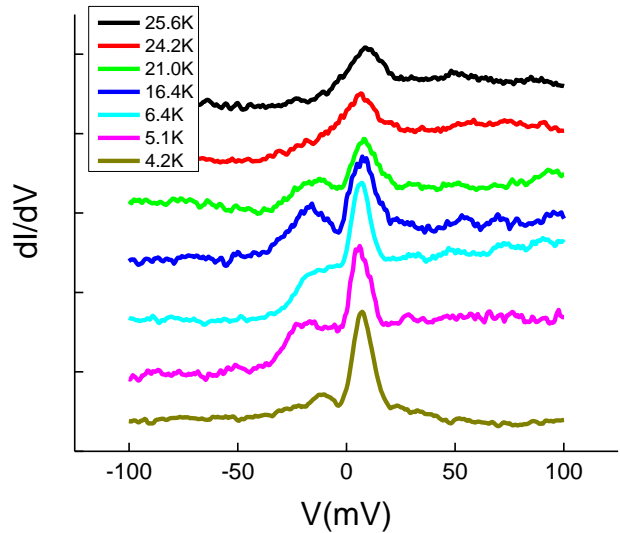

**Fig.2S**. Temperature dependence of the Kondo peaks. V = -200mV, I = 20pA, Vg = 50V.

## 5. Substrate effect on the Kondo coupling.

As discussed in the main text and consistent with the theoretical predictions, Kondo screening is observed for the rougher $G/SiO_2$ and $G/G/SiO_2$ samples while it is absent on the much smoother G/BN and G/G/BN. Interestingly, most of the earlier work focused on the in-plane Jahn-Teller distortion but not on the out-of plane distortion. This is the first work to show the crucial role of the local curvature. Fig.3S shows the Kondo peak for a vacancy in the G/SiO$_2$ sample where the local curvature is largest. $T_K$ here is much higher than that on G/G/SiO$_2$ consistent with the rougher surface.

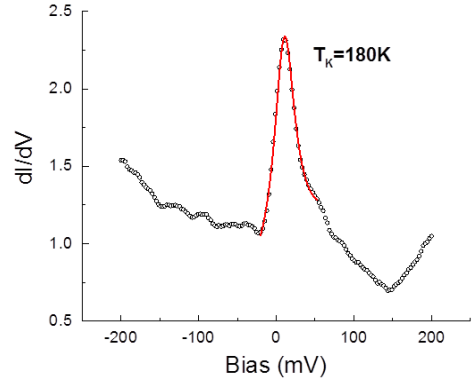

**Fig.3S.** dI/dV curve together with Fano-fit (red) for a vacancy in G/SiO$_2$, $V_b$=-300mV, I = 20pA.

## 6. Criteria to identify the intrinsic single vacancy

As discussed in the main text, several criteria were used to identify the intrinsic single atom vacancies.

Firstly, theoretical and experimental work has shown that single atom vacancies display a characteristic triangular structure arising from the electronic reconstruction (*11, 12*). This triangular fingerprint is used to distinguish between bare vacancies and vacancies that are passivated by adsorbed adatoms such as hydrogen or nitrogen whose topography lacks this feature (*13, 14*). Secondly, the zero mode peak at the Dirac point allows to distinguish between single-atom vacancies and other types of un-passivated defects, such as Stone-wales reconstruction or di-vacancies, which lack both the triangular electronic reconstruction feature and the zero mode in the electronic states (*4, 15*). Thirdly, the evolution of the local Dirac point with the distance from the single vacancy center is examined by the dI/dV curves. If a chemical bond forms between the dangling bond and an ad-atom, the electron affinity difference will cause a charge transfer at the vacancy site. This shifts the local Dirac point and gives rise to a clear shift of the Dirac across the vacancy site (*16*). Combining all these criteria together ensures the correct identification of single atom vacancies.

## 7. Pseudogap Anderson impurity model

The Kondo effect for single carbon vacancies in graphene is described by the pseudogap single-channel asymmetric Anderson impurity model(*17*). The corresponding Hamiltonian can be written as(*18*):

$$H = \sum_\sigma \int_{-D}^{D} d\omega \frac{(\omega+\mu)}{D} c_\sigma^\dagger(\omega) c_\sigma(\omega) + \sum_\sigma (\varepsilon_d - \mu) f_\sigma^\dagger f_\sigma + U_{eff}(\mu) f_\uparrow^\dagger f_\uparrow f_\downarrow^\dagger f_\downarrow +$$

$$\sum_\sigma \int_{-D}^{D} d\omega \sqrt{\frac{\Gamma(\omega)}{\pi D}} [c_\sigma^\dagger(\omega) f_\sigma + f_\sigma^\dagger c_\sigma(\omega)],$$

where spin index $\sigma = \uparrow, \downarrow$; $D$ is the conduction bandwidth; $c_\sigma^\dagger(\omega)[c_\sigma(\omega)]$ is the creation (annihilation) operator for an electron in the conduction state with energy $\omega$; $\mu$ is the chemical potential; $\varepsilon_d$ is the energy of the impurity level; $U$ is the Coulomb interaction between the electrons on the impurity; $\Gamma(\omega)$ is the coupling function; $f_\uparrow^\dagger$ ($f_\downarrow^\dagger$) and $f_\uparrow$ ($f_\downarrow$) are creation and annihilation operators for an electron in the $\uparrow$ ($\downarrow$) impurity state.

Since the zero-mode (ZM) is near the Dirac point [see, e.g., (*8*, *16*, *17*) and also Fig. 1D and Fig. 1E in the main text] it does not contribute to the Kondo critical behavior in the p-doped regime ($\mu < 0$) as long as the ZM is unoccupied. In the n-doped regime ($\mu > 0$) however when the ZM becomes occupied, the energy of the doubly occupied σ-orbital could rise above the onsite Coulomb interaction $U$ of the σ-orbital due to the additional Coulomb repulsion from the ZM. This increases the effective Coulomb interaction of the σ-orbital state. Therefore, we take this into account by adopting the following effective Coulomb interaction between the electrons on the impurity:

$$U_{eff}(\mu) = \begin{cases} U & \mu \leq 0 \\ U + \min(U_{\pi d}, \alpha\mu) & \mu > 0 \end{cases}$$

where $U_{\pi d}$ is the Coulomb interaction between the ZM and σ-orbital; $\alpha$ is a positive constant.

The coupling function can be written as [8]:

$$\Gamma(\omega) = \frac{\Omega_0 V^2 |\omega + \mu|}{2\hbar^2 v_F^2}\left[2 - J_0\left(\frac{2}{3}\frac{|\omega + \mu|}{t}\right)\right],$$

where $\Omega_0, V, t,$ and $v_F$ are the unit cell area, the hybridization strength, hopping energy, and Fermi velocity, respectively. We can expand the zeroth Bessel function $J_0$ at $\omega = 0$, and $\Gamma(\omega)$ can be approximated as:

$$\Gamma(\omega) = \frac{\Omega_0 V^2 |\omega + \mu|}{2\hbar^2 v_F^2}\left[1 + \frac{4}{27}\left(\frac{|\omega + \mu|}{t}\right)^2\right].$$

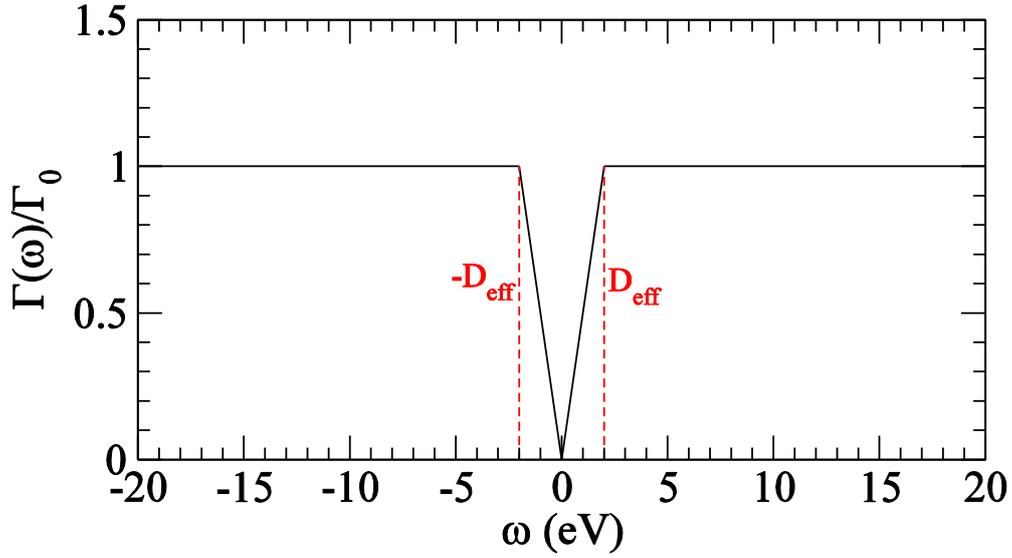

**Fig. 4S**. The coupling function $\Gamma(\omega)$ adopted in this work. The total bandwidth $D = 20$ eV and effective bandwidth $D_{\text{eff}} = 2$ eV are taken from the graphene band structure from ab initio density functional calculations.

Earlier, we showed that the second term hardly affects the calculated spectral function(*19*), and thus we approximate the $\Gamma(\omega)$ as a linear function of $\omega$. In this work, we use the mixed form in which $\Gamma(\omega)$ is proportional to $|\omega + \mu|$ only within the effective bandwidth $D_{eff}$, and is constant for the rest of the bandwidth (see Fig. 4S). Mathematically, $\Gamma(\omega)$ can be written as:

$$\Gamma(\omega) = \begin{cases} \Gamma_0 \dfrac{|\omega + \mu|}{D_{eff}} & \dfrac{|\omega + \mu|}{D_{eff}} \leq 1 \\ \Gamma_0 & \dfrac{|\omega + \mu|}{D_{eff}} > 1, \dfrac{|\omega|}{D_{eff}} \leq \dfrac{D}{D_{eff}} \\ 0 & \dfrac{|\omega|}{D_{eff}} > \dfrac{D}{D_{eff}} \end{cases},$$

where $\Gamma_0 = \dfrac{\Omega_0 V^2 D_{eff}}{2\hbar^2 v_F^2}$ represents the coupling strength discussed in the main text.

## 8. Numerical renormalization group calculations

We exploit the powerful numerical renormalization group (NRG) method to solve the Anderson impurity model(20). In the present calculations, we use the discretization parameter $\Lambda = 1.8$ and keep 1200 states per NRG iteration so that the obtained spectral functions converge within 0.1 %. We set the total bandwidth $D = 20$ eV and $D_{eff} = 2$ eV (19) (Fig. 4S). Note that the numerical results do not depend on $D$ and $D_{eff}$ as long as they are larger than 8 eV and 1 eV, respectively. Typically, the onsite Coulomb interaction varies from about 1 to 10 eV. Our experimental results indicate that the singly occupied impurity state ($\varepsilon_d$) is well below the Fermi level and only the doubly occupied impurity state ( $U + \varepsilon_d$ ) is relevant here (i.e., $|U + \varepsilon_d| < |\varepsilon_d|$). Starting with the LDA value $U = 2$ eV for the Coulomb interaction strength and $\varepsilon_d = -1.5$ eV for the bare σ-orbital energy (19, 21), we fit several theoretical spectral functions for slightly varied $U$ and $\varepsilon_d$ values to a typical experimental dI/dV curve, and we got the impurity level $\varepsilon_d = -1.6$ eV and $U = 2$ eV. With this set of the model parameters, we performed the NRG calculations of the spectral function for several values of $\Gamma_0$ in order to determine the critical value $\Gamma_C$ which separates the local-moment regime and the frozen-impurity regime at $\mu = 0$. We found $\Gamma_C = 1.15$ eV from the results of these NRG calculations. Finally, by comparing the sets of spectral functions (Fig. 5s), we set $\alpha=3.5$ and $U_{\pi d} = 0.2$ eV.

We estimated the Kondo temperatures by fitting the spectral functions calculated at the experimental temperature T = 4.2 K to the Fano lineshape, which allows us to draw the phase diagram shown in Fig. 4C. Figure 4B shows the Kondo temperature versus the chemical potential. For $\Gamma_0 < \Gamma_C$, Kondo screening occurs in both p-doped and n-doped sectors. The NRG simulations of the spectra in Fig. 5s A, indicate a value of $\Gamma_0 = 0.90\Gamma_C$ for this vacancy, corresponding to the subcritical regime. For $\Gamma_0 > \Gamma_C$, the system flows to the FI regime at negative $\mu$. For the spectra in Fig. 5s B, we find $\Gamma_0 = 1.83\Gamma_C$ which puts this vacancy in the supercritical regime. A comparison between the measured (Fig 3A) and simulated (Fig3B) $T_K(\mu)$ curves was used to obtain the $\Gamma_0/\Gamma_C$ values characterizing each vacancy.

## 9. Simulated gate dependent dI/dV curves for different vacancies

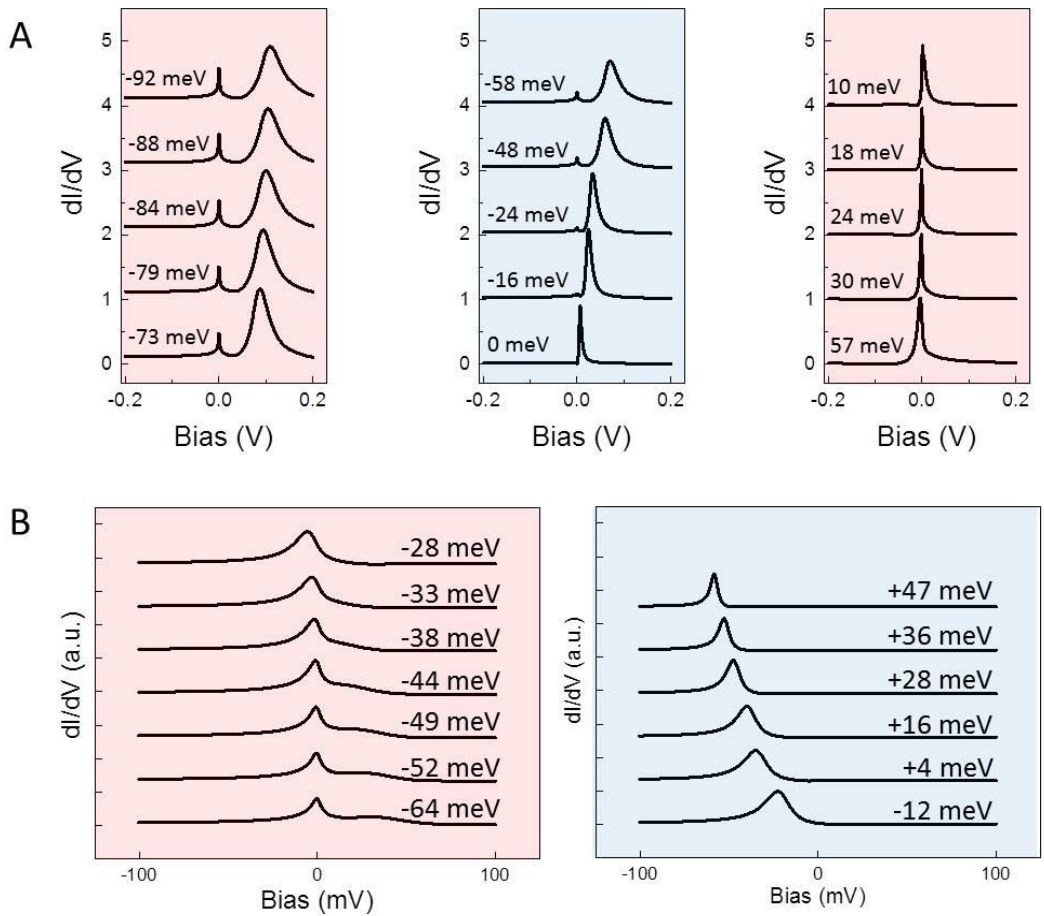

Fig.5s shows the NRG simulation of the experimental spectra presented in Fig.2.

References

1. A. Luican, G. Li, E. Y. Andrei, Scanning tunneling microscopy and spectroscopy of graphene layers on graphite. *Solid State Communications* **149**, 1151 (2009).

2. G. Li, A. Luican-Mayer, D. Abanin, L. Levitov, E. Y. Andrei, Evolution of Landau levels into edge states in graphene. *Nat Commun* **4**, 1744 (2013).
3. C.-P. Lu *et al.*, Local, global, and nonlinear screening in twisted double-layer graphene. *Proceedings of the National Academy of Sciences*, (June 2, 2016, 2016).
4. J. Mao *et al.*, Realization of a tunable artificial atom at a supercritically charged vacancy in graphene. *Nat Phys* **12**, 545 (2016).
5. G. Li, A. Luican, E. Y. Andrei, Self-navigation of an STM tip toward a micron sized sample. *Rev. Sci. Instrum.* **82**, (2011).
6. E. Y. Andrei, G. H. Li, X. Du, Electronic properties of graphene: a perspective from scanning tunneling microscopy and magnetotransport. *Rep Prog Phys* **75**, 056501 (May, 2012).
7. A. Luican-Mayer *et al.*, Screening Charged Impurities and Lifting the Orbital Degeneracy in Graphene by Populating Landau Levels. *Phys Rev Lett* **112**, 036804 (2014).
8. A. Luican *et al.*, Single-Layer Behavior and Its Breakdown in Twisted Graphene Layers. *Phys Rev Lett* **106**, 126802 (2011).
9. U. Fano, Effects of Configuration Interaction on Intensities and Phase Shifts. *Physical Review* **124**, 1866 (1961).
10. V. Madhavan, W. Chen, T. Jamneala, M. F. Crommie, N. S. Wingreen, Tunneling into a Single Magnetic Atom: Spectroscopic Evidence of the Kondo Resonance. *Science* **280**, 567 (1998-04-24 00:00:00, 1998).
11. H. Amara, S. Latil, V. Meunier, P. Lambin, J. C. Charlier, Scanning tunneling microscopy fingerprints of point defects in graphene: A theoretical prediction. *Phys Rev B* **76**, 115423 (2007).
12. F. Banhart, J. Kotakoski, A. V. Krasheninnikov, Structural Defects in Graphene. *ACS Nano* **5**, 26 (2011/01/25, 2011).
13. L. Zhao *et al.*, Visualizing Individual Nitrogen Dopants in Monolayer Graphene. *Science* **333**, 999 (2011).
14. M. Ziatdinov *et al.*, Direct imaging of monovacancy-hydrogen complexes in a single graphitic layer. *Phys Rev B* **89**, 155405 (2014).
15. E. Zaminpayma, M. E. Razavi, P. Nayebi, Electronic properties of graphene with single vacancy and Stone-Wales defects. *Applied Surface Science* **414**, 101 (2017/08/31/, 2017).
16. C. Ma *et al.*, Tuning the Doping Types in Graphene Sheets by N Monoelement. *Nano Letters*, (2017/12/21, 2017).
17. C. Gonzalez-Buxton, K. Ingersent, Renormalization-group study of Anderson and Kondo impurities in gapless Fermi systems. *Phys Rev B* **57**, 14254 (1998).
18. T. Kanao, H. Matsuura, M. Ogata, Theory of Defect-Induced Kondo Effect in Graphene: Numerical Renormalization Group Study. *J Phys Soc Jpn* **81**, 063709 (2012).
19. P. W. Lo, G. Y. Guo, F. B. Anders, Gate-tunable Kondo resistivity and dephasing rate in graphene studied by numerical renormalization group calculations. *Phys Rev B* **89**, 195424 (2014).

20. K. G. Wilson, The renormalization group: Critical phenomena and the Kondo problem. *Rev Mod Phys* **47**, 773 (1975).
21. O. V. Yazyev, L. Helm, Defect-induced magnetism in graphene. *Phys Rev B* **75**, 125408 (2007).